\documentstyle[11pt,newpasp,twoside,epsf,psfig]{article}
\def\edcomment#1{\iffalse\marginpar{\raggedright\sl#1\/}\else\relax\fi}
\reversemarginpar

\begin{document}
\title{Sinks of Light Elements in Stars -- Part III}
 \author{Corinne Charbonnel}
\affil{Laboratoire d'Astrophysique de l'Observatoire Midi-Pyr\'en\'ees, CNRS-UMR
5572, 14, av.E.Belin, F-31400 Toulouse, France}
\author{Constantine P. Deliyannis}
\affil{Indiana University, Astronomy Department,
319 Swain Hall West, 727 E. 3rd Street, Bloomington, IN 47405-7105, USA}
\author{Marc Pinsonneault}
\affil{The Ohio State University, Department of Astronomy, 140
W. 18th Ave., Columbus, OH 43210, USA}

\begin{abstract}
See the abstract given in Part I (Deliyannis, Pinsonneault, Charbonnel, 
hereafter DPC, this volume).
In Part III, we first discuss the LiBeB observations in subgiant stars and the 
constraints 
they bring on the transport processes occuring on the main sequence. 
Evidence is then reviewed that suggests that in situ mixing occurs in
evolved low mass Population I and Population II stars. Theoretical
mechanisms that can create such mixing are discussed, as well as their
implications for the evolution of the LiBeB and $^3$He.
\end{abstract}

\section{Introduction}
Evolved stars can be important sinks for light elements.  In Parts I and II we
reviewed the observational and theoretical situation for Population I and
Population II main sequence stars. The structural evolution of evolved low mass
stars is different than their main sequence precursors. 
The surface abundances are diluted as stars travel from the main
sequence across the subgiant branch to the red giant branch.  This first
dredge-up is completed when the surface convection zone reaches its maximum
depth in mass (below the luminosity of the horizontal branch).  The properties 
of stars as they undergo the first dredge-up can provide clues about the internal
abundance profiles of their main sequence precursors; we review the conclusions
that can be drawn from subgiants in \S 2.  On the upper giant branch standard
models - which predict constant surface abundances after the completion of the 
first
dregdge-up - are in serious conflict with the observational data, which exhibits 
strong
trends with increased luminosity; we discuss first-ascent red giants in \S 3.

\section{LiBeB on the subgiant branch -
Constraints for the processes occuring on the main sequence}

\subsection{Population I subgiants}

LiBeB abundances in subgiants are {\sl a posteriori} tracers of the 
hydrodynamical processes that affect these elements 
during the previous evolutionary phases.
Indeed, when the first dredge-up occurs, the convective dilution of the external 
stellar layers with the internal LiBeB free regions induces a decrease of the 
surface abundances down to values that depend on the stellar mass and 
metallicity 
(which dictate the dredge-up efficiency) and on the total LiBeB content in the
star at the turnoff.
One thus expects very low post-dilution abundances for stars which significantly 
destroyed these elements during the pre-main sequence and the main sequence.  
This is the case in particular for Li in Pop I evolved stars with initial masses 
lower than about 1.4M$_{\odot}$ for which the large Li dispersion 
reflects the distribution on the main sequence and at the turnoff (see DPC).

Until recently, the situation was less clear for the more massive stars 
(A and early-F types) which spend their main sequence on the hot side 
of the Li-dip (note that there are strong observational selection effects for 
the 
hotter stars which are rapidly rotating and for which lithium abundances can not 
be 
measured). 
In open clusters like Coma, Praesepe and the Hyades, these stars show Li 
abundances 
close to the galactic value, except for some Li deficient Am stars (Boesgaard 
1987; 
Burkhart \& Coupry 1989, 1998, 2000).
However, important Li underabundances are exhibited by some of their field main 
sequence or slightly evolved counterparts before the dilution starts 
(Alschuler 1975, Brown et al. 1989, Balachandran 1990, Wallerstein et al. 1994). 
This, in addition to the fact that dilution alone can not
explain the low lithium values shown by the giants in the open clusters
with turnoff masses higher than $\sim$ 1.5M$_{\odot}$ (Gilroy 1989, Charbonneau
et al. 1989; see Figure 1) 
suggested that some Li depletion occurs inside these stars while they are 
on the main sequence but shows up at the surface relatively late 
(i.e., still on the main sequence but after the age of the Hyades) 
compared to cooler dwarfs (Vauclair 1991, Charbonnel \& Vauclair 1992).
This was confirmed by Randich et al. (1999) and do Nascimento et al. (2000,
using data by L\`ebre et al. 1999) on the basis of the spectroscopic analysis 
of large samples of field Pop I subgiants for which Hipparcos data allowed 
the precise determination of both the mass and evolutionary status (see also 
Mallik 1999 and in this volume)\footnote{Let us note that the very low Li 
values found for the most massive subgiants are confirmed even when 
non-LTE effects are taken into account (do Nascimento et al. 2000)}. 
Observations of Be and B in a few Hyades giants brought additional constraints 
on the
processes that occur in the external layers of the A and early-F main sequence 
stars : 
In these giants indeed, Be is moderately underabundant (Boesgaard et al. 1977) 
while B is almost normal (Duncan et al. 1998) compared to the dilution predictions.

\begin{figure}
\centerline{
\psfig{figure=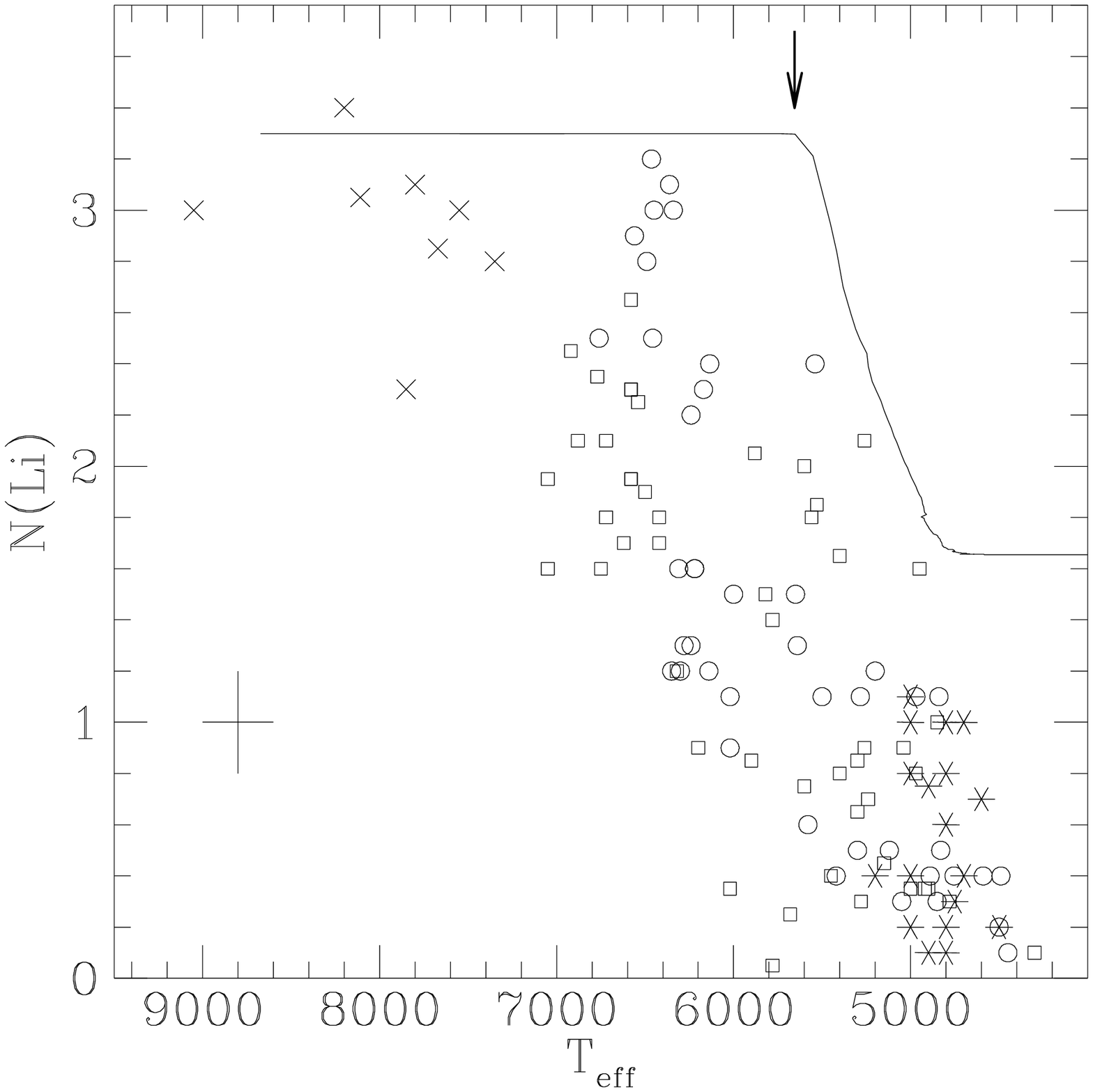,height=5cm}
\psfig{figure=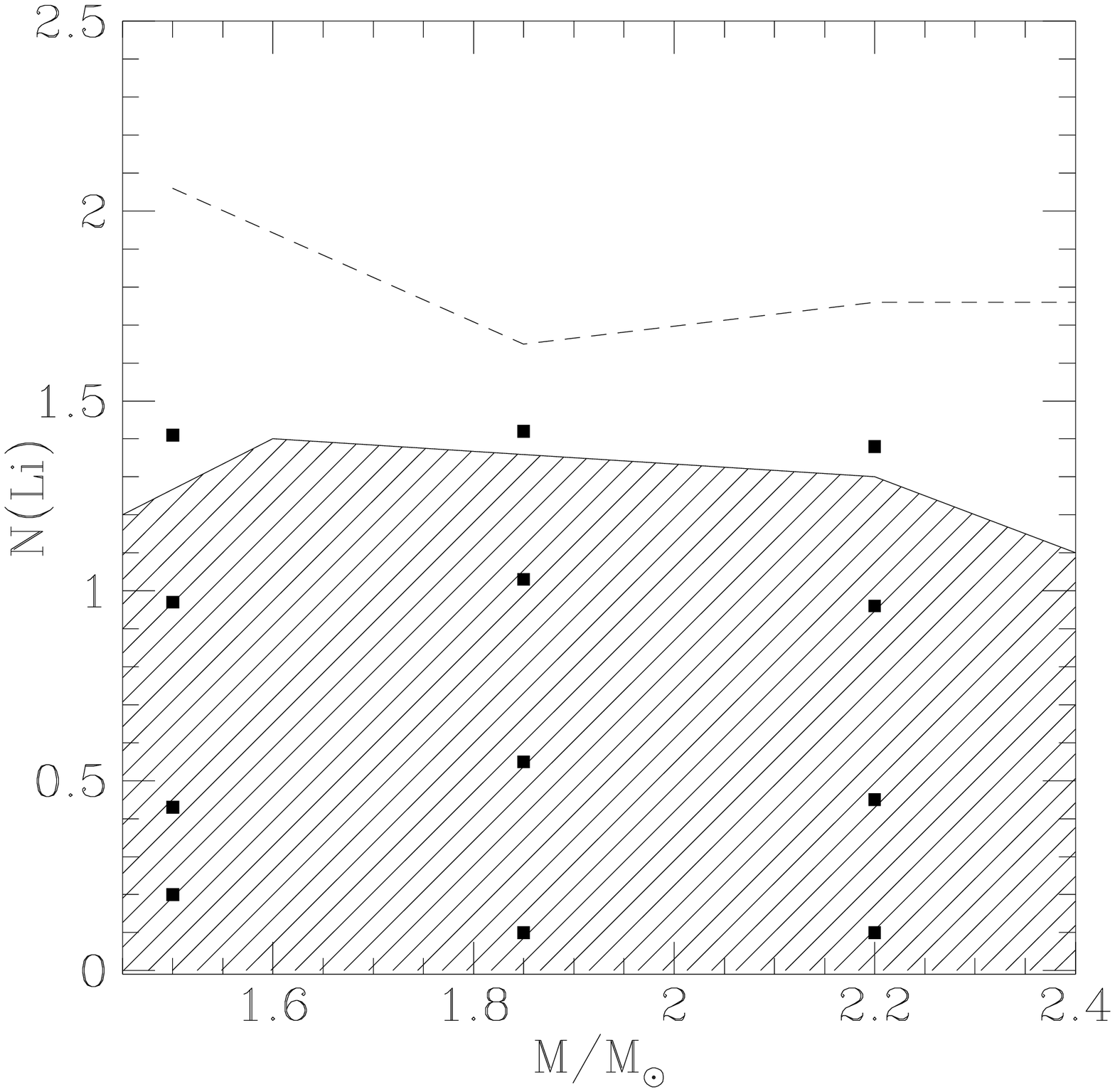,height=5cm} }
\caption{Li in stars with masses higher than 1.5M$_{\odot}$. 
{\bf (left)} Hyades main sequence stars (crosses, Burkhart \& Coupry 2000), 
field subgiants (circles, L\`ebre et al. 1999; squares, Wallerstein
et al. 1994), giants of open clusters (stars, Gilroy 1989). 
The solid line shows the evolution of the surface lithium with 
T$_{\rm eff}$ for a standard 1.85M$_{\odot}$ model (Charbonnel \& Talon
1999, CT99); the arrow points the theoretical start of dilution.
{\bf (right)} Evolved stars of open clusters (shaded area).
The dotted curve shows the Li values expected with dilution alone, 
i.e. no additional internal mixing on the main sequence. 
The squares give the Li values predicted when rotation induced mixing 
is taken into account as in CT99 for different 
rotation velocities of the main sequence star ranging 
between 50 and 150 km.sec$^{-1}$
%(An initial value of N(Li)=3.3 is assumed)
}
\end{figure}

In A and F main sequence stars, atomic diffusion is known to play an 
important role (mainly in the very outermost stellar layers; see Michaud in this 
volume). 
However, diffusion theory alone predicts abundance anomalies much larger than 
the 
observed ones (Richer et al. 2000 and references therein). 
In particular at the age of the Hyades, radiative
diffusion in stars with T$_{\rm eff} \simeq$7000K should lead to important Li
overabundances, while at T$_{\rm eff} >$7200K strong Li underabundances should 
be
due to settling. This is not observed. 
Some macroscopic processes thus occur, that both decrease the efficiency 
of the atomic diffusion and lead to non standard Li depletion in evolved stars
as we discussed previously 
(see also Part II, Pinsonneault, Charbonnel \& Deliyannis in this volume)

Charbonnel \& Talon (1999, hereafter CT99) studied the combined
effect of atomic diffusion and rotational mixing on the LiBeB in these stars up 
to the completion of the first dredge-up, in the framework of the transport of 
matter and of angular momentum by wind-driven meridional circulation and 
shear turbulence (Zahn 1992, Maeder 1995, Talon \& Zahn 1997, Maeder \& Zahn 1998).
Their models were computed for rotation velocities covering the 
large Vsini range observed at these spectral types.
While lithium is found not to vary much at the stellar surface 
at the age of the Hyades, more destruction occurs inside the 
rotating models compared to the classical ones and 
its signature appears at the surface before the onset of the dilution. 
The post dredge-up Li values are much lower than predicted classically and
agree with the observations both in evolved stars belonging to open clusters
and in the field, as can be seen in Fig.1.
The less fragile Be and B are less affected than Li by the rotation-induced mixing, 
and the corresponding predictions also reproduce the observations in the Hyades 
giants (Boesgaard et al. 1977, Duncan et al. 1998).
The main success of CT99 models which include the most complete description
currently available for rotation-induced mixing is their ability to reproduce 
abundance anomalies of various elements over a large domain of stellar masses 
and evolutionary phases. 
Indeed the same treatment of the hydrodynamical process which can account for 
the C and N anomalies in B type stars (Talon et al. 1997) 
also shapes the hot side of the Li dip in the open clusters 
(Talon \& Charbonnel 1998) and explains the LiBeB observations in 
main sequence F and A stars as well as in their evolved counterparts 
(CT99; see also Talon \& Charbonnel in this volume).

\subsection{Population II subgiants}
 
The behavior of lithium in metal-poor field subgiants appeared clearly 
in the large sample of Pilachowski et al. (1993) and was confirmed 
recently by Gratton et al. (2000; see also Carretta et al. 1998). 
%[Fe/H] between about -2 and -1.
As can be seen in Fig.2, 
the observed trend (steady decline of lithium abundance with temperature
decreasing between $\sim$ 5600 and 4900K on the subgiant branch) 
is well explained by the theoretical dilution up to 
the completion of the first dredge-up (e.g. Deliyannis et al. 1990, 
Proffitt \& Michaud 1991, Charbonnel 1995). 
The precise temperature at which theoretical dilution begins is somewhat 
model dependent, but nonetheless, given current lingering uncertainties 
in the temperature scales, the agreement between theory and observation 
is impressive.

Pilachowski et al. (1993) noted that the lithium abundances in their subgiant 
sample showed more scatter than do the turnoff stars; they interpreted 
this as an indication for variations in the main sequence Li destruction 
below the observable surface layers. 
Reconsidering Pilachowski's sample to which they added some more stars,
Ryan \& Deliyannis (1995) showed however that much of the scatter disappears 
with a careful and self-consistent treatment of the reddening. 
Note that this difference does not show up in Gratton's sample.
However, the PopII field subgiants with Li upper limits
are presumably the counterparts of the few plateau stars with no Li detection
(see Part II in this volume); 
both samples should be analysed simultaneously to understand and
quantify a possible Li depletion in halo stars. 

\begin{figure}
\centerline{ 
\psfig{figure=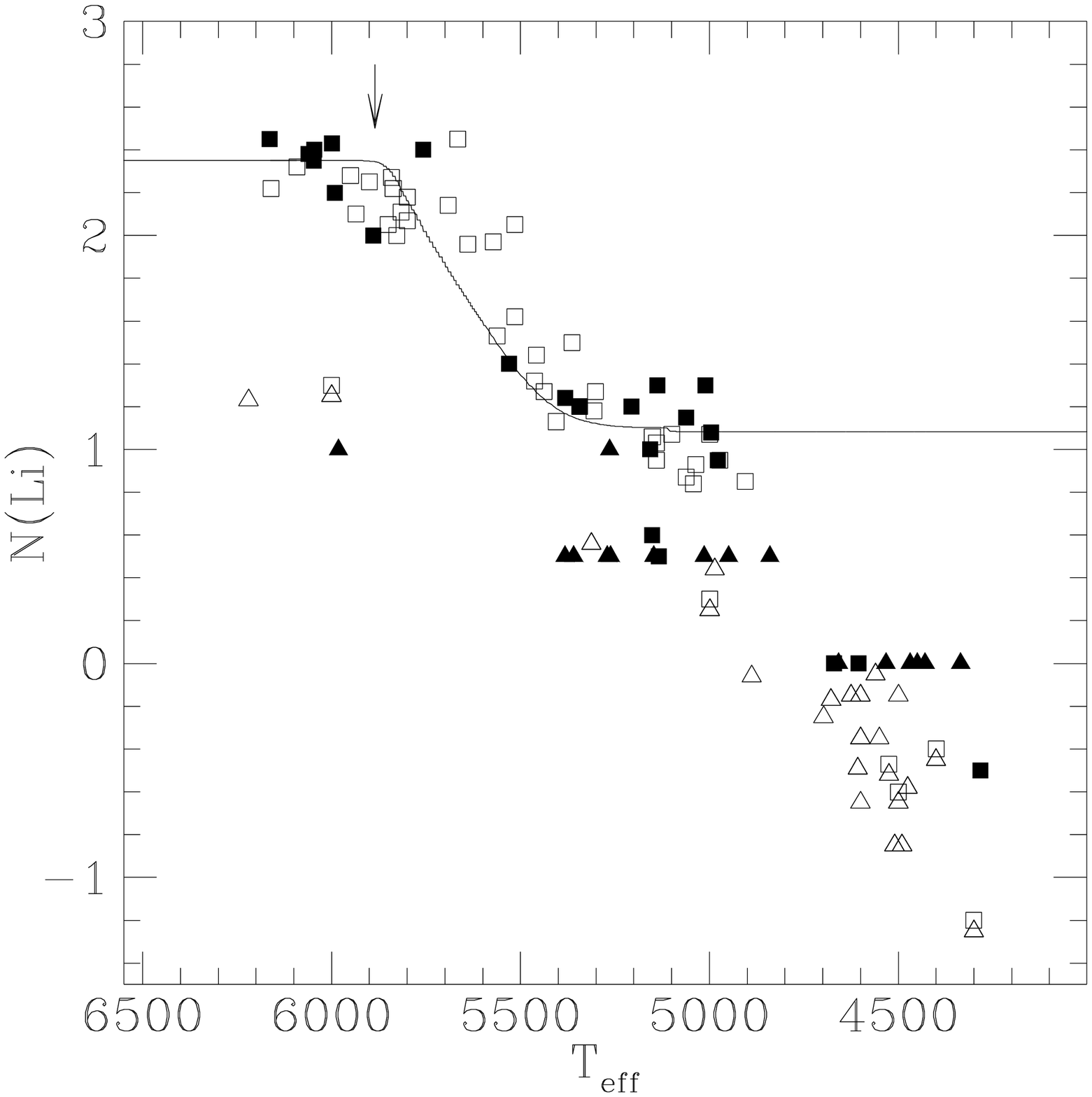,height=5.5cm}
\psfig{figure=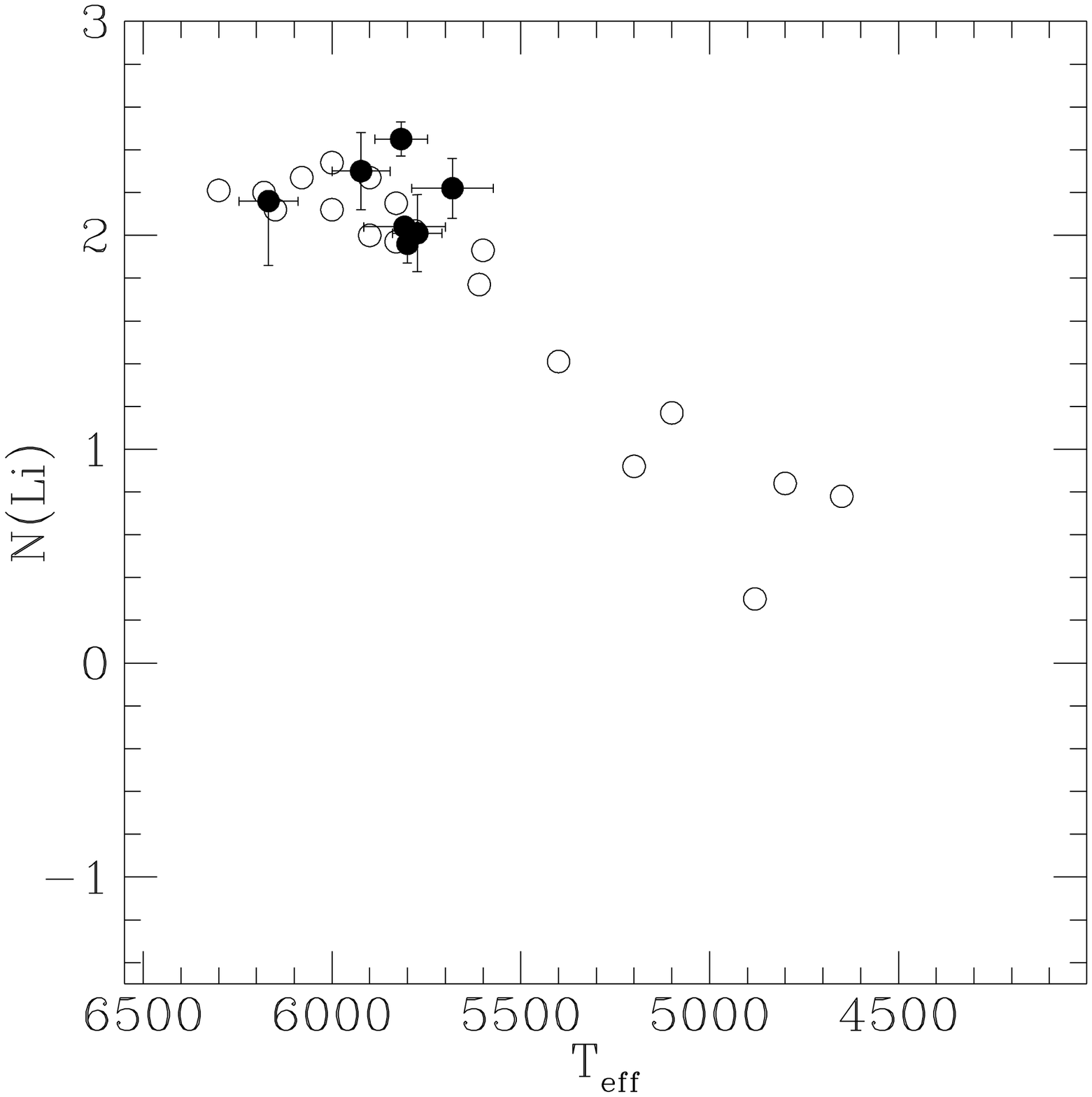,height=5.5cm} }
\caption{Li in metal-poor subgiants. 
{\bf (left)} Field stars with $-2\leq$[Fe/H]$\leq-1$ (Ryan \& Deliyannis 1995, 
open squares and triangles for real detection and upper limits respectively; 
Gratton et al. 2000, black squares and triangles).
The solid line shows the prediction of the Li variations due to dilution 
alone (which starts as indicated by the arrow) 
at the surface of a 0.8M$_{\odot}$, Z=10$^{-4}$ star (Palacios et al. 2000).
{\bf (right)} Stars in globular clusters M92 
(Deliyannis et al. 1995, Boesgaard et al. 1998, black points) 
and NGC 6397 (Molaro \& Pasquini 1994, Pasquini \& Molaro 1996); 
the values presented here were derived within Carney's temperature scale 
(Spite, private com.)} 
\end{figure}

Very few data exist up to now for LiBeB in globular clusters. 
Lithium abundances have been determined for slightly evolved stars 
which have not reached the onset of dilution 
in NGC 6397 (Molaro \& Pasquini 1994, Pasquini \& Molaro 1996) 
and M92\footnote{The globular cluster stars were chosen to be pre-dilution
stars, relative to the empirical commencement of the Li dilution from
field stars} (Deliyannis et al.  1995, Boesgaard et al. 1998). 
These turnoff stars with very similar effective temperature show a Li dispersion
of a factor about two and three respectively in NGC 6397 and M92 (see Fig.2). 
Boesgaard et al.(1998) favor the explanation of differential depletion in 
M92 by rotation induced mixing in objects with different stellar 
rotational histories.
It is worth knowing however that the M92 subgiants show other ``surprising
abundances" (King et al. 1998), i.e., under and overabundances respectively 
of Mg and Na compared to field stars with the same metallicity. 
While the observational data need to be confirmed, these anomalies reveal 
stricking field to cluster differences (see also \S 3) which could 
reflect environmental effects (pollution of the intracluster gas, 
distribution of the initial rotation velocities, ...) which have to be 
disentangled. This is of particular importance if one wants to use and
compare the Li data in globular cluster and halo stars to constrain 
the primordial abundance of this element and its evolution in the 
early epochs (see Part II).

\section{LiBeB and $^3$He on the Red Giant Branch}

\subsection{Evidences for extra-mixing in low-mass RGB stars}

Observational evidences have accumulated during the last years 
of a second and distinct mixing episode that occurs in low mass 
stars when they climb the red giant branch (RGB; 
see Kraft 1994, and more recently Charbonnel et al. 1998, Sneden 1999 
and Gratton et al. 2000 for references). 
The signatures of this non-standard process in terms of abundance 
anomalies are numerous. 
In metal-poor field giants, it leads to a further major decrease of 
the Li abundance (around 4800K as can be seen in Fig.2). 
By reaching the regions of incomplete CNO burning inside the RGB stars, 
it induces a decrease of the carbon abundance and of the carbon 
isotopic ratio, and a corresponding increase of the N. In most of the 
metal-deficient field and globular cluster stars, the surface 
$^{12}$C/$^{13}$C ratio even approaches the equilibrium value; 
this anomaly also appears, while at a lower extent, in evolved stars 
belonging to open clusters with turnoff masses lower than $\sim$2M$_{\odot}$
(Gilroy 1989, Gilroy \& Brown 1991). 
This extra-mixing is also frequently invoqued to explain the global 
O versus Na anticorrelation observed in globular cluster red giants
(see Weiss et al. 2000 for references).

All the relevant data clearly indicate that the extra-mixing starts 
acting when the stars pass the so-called RGB bump in the luminosity function.
At this point, the hydrogen burning shell (HBS) crosses the discontinuity 
in molecular weight built by the convective envelope during the 
dredge-up.
Before this evolutionary point, the mean molecular weight gradient 
probably acts as a barrier to the mixing between the convective envelope
and the HBS 
(Sweigart \& Mengel 1979, Charbonnel 1994, 1995, 
Deliyannis 1995, Charbonnel et al. 1998). 
Above this point, this barrier disappears and the extra-mixing, 
whatever its nature, is free to act.  

\subsection{The origin of the extra-mixing ...}

Several attempts have been made to simulate this extra-mixing in
order to reproduce the abundance anomalies in RGB stars.
Denissenkov \& Weiss (1995) and Weiss et al. (2000) 
modelled this deep mixing by adjusting both the mixing depth and rate 
in their diffusion procedure, but focussed on the O-Na anticorrelation
(see also Cavallo et al. 1998). 
Wasserburg et al. (1995), Boothroyd  \& Sackmann (1999) and Sackmann \& 
Boothroyd 
(1999; see also Sackmann in this volume) used an ad-hoc ``conveyor-belt" 
circulation model, where the depth of the ``extra-mixed region" is related 
to a parametrized temperature difference up to the bottom of the HBS and 
is adjusted to reach the observed carbon isotopic ratios as a function of 
stellar mass and metallicity. 

Some studies attempted to relate the RGB extra-mixing with physical processes, 
among which rotation seems to be the most promising.
Sweigart \& Mengel (1979) suggested that meridional circulation on
the RGB could lead to the low $^{12}$C/$^{13}$C observed in field giants.
Charbonnel (1995) investigated the influence of such a process 
by taking into account more recent progress in the
description of the transport of chemicals and angular momentum in
stellar interiors : Zahn's (1992 and subsequent developments) 
consistent theory which describes the interaction between 
meridional circulation and turbulence induced by rotation 
(as already discussed in \S 2.1). 
This framework is appealing because it takes advantage of some particularities 
of the non-homologous RGB evolution. 
In particular, some mixing is expected to take place wherever the rotation 
profile presents steep vertical gradients, and near nuclear burning shells.
Moreover due to the stabilizing effect of the composition gradients, 
the mixing is expected to be efficient on the RGB only when the
hydrogen-burning-shell crosses the chemical discontinuity
created by the convective envelope during the first dredge-up.
Using a simplified version of this description Charbonnel (1995) showed that 
the rotation-induced mixing can indeed account for the observed behavior 
of carbon isotopic ratios and for the Li abundances in Population II low mass 
giants. 
Simultaneously, when this extra-mixing begins to act, $^3$He is
rapidly transported down to the regions where it burns by the
$^3$He($\alpha, \gamma)^7$Be reaction. This leads to a decrease
of the surface value of $^3$He/H (see also Deliyannis 1995, Hogan 1995 and 
Sackmann \& Boothroyd 1999).

The study of horizontal branch stars also provides some intriguing
clues about angular momentum evolution on the RGB.  Peterson (1983)
first discovered that some blue horizontal branch stars are rapid
rotators (see Behr et al. 2000 for more recent data and a discussion
of the observational situation).  Pinsonneault et al. (1991) noted that
the combination of RGB mass loss, high horizontal branch rotation, and
low main sequence rotation required strong differential rotation with
depth in giants.  If the convection zone of RGB stars had solid body
rotation, differential rotation with depth in their MS precursors was
also required.
Behr et al. (2000) found a break in the rotational properties of HB
stars, in the sense that very blue HB stars both exhibited the surface
signature of atomic diffusion and rotated more slowly than slightly
cooler stars.  Sills \& Pinsonneault (2000) interpreted this as
evidence that mean molecular weight gradients caused by atomic
diffusion inhibit angular momentum transport in hot horizontal branch
stars; this is an independent test of the impact of composition
gradients in a different evolutionary phase.  In addition, Sills \&
Pinsonneault (2000) found that uniform rotation at the main sequence turnoff
was only compatible with rapid horizontal branch rotation under the
following conditions: 
(1)turnoff rotation of order 4 km/s rather than the 1 km/s inferred
from an extrapolation of the Population I angular momentum loss law to
Population II stars; 
(2)constant specific angular momentum in the convective envelopes of
giants; 
(3)strong differential rotation with depth in the radiative cores of
giants.
All of these are radically different from the expectations from main
sequence angular momentum evolution models, and they are an indication
that further theoretical work is needed in physical models of RGB
rotational mixing.  It is encouraging, however, that all of the above
properties favor more vigorous rotational mixing on the RGB than would
be expected from the opposite conclusions.

\subsection{... and its consequences for $^3$He}

While there is a consensus on the fact that 
the mechanism which is responsible for the chemical anomalies on the RGB
also affects the $^3$He abundance (as first suggested by Rood et al.
1984), large uncertainties remain on the quantitative extent of this 
$^3$He depletion. In the PopII models of Charbonnel (1995) $^3$He
decreases by a large factor in the ejected envelope material but low
mass stars remain net producers (while far much less efficient than in the 
case of models without RGB extra-mixing) of $^3$He. 
On the other hand the models of Sackmann \& Boothroyd (1999) predict a
net destruction of $^3$He by low mass stars. 
Both studies also differ on the predictions for the evolution of the 
$^{12}$C/$^{13}$C ratio at the stellar surface. While the
rotation-induced mixing by Charbonnel reproduces the observed sharp drop 
of the $^{12}$C/$^{13}$C just beyond the RGB bump and its constancy at 
higher luminosity, the cool bottom processing model by Sackmann \& Boothroyd 
predicts a smooth decrease of the surface $^{12}$C/$^{13}$C all along 
the RGB up to the tip where the low values are finally (but too belatedly 
compared to the data) reached. 

As is customary, we can say that at this stage additional studies are 
needed. In particular, a consistent treatment of the transport of 
chemicals and angular momentum as decribed in \S 2.1 will prove most useful
to quantify with confidence the impact of the rotation-induced mixing on
the RGB. Work is in progress in this direction (Palacios et al. 2000).

The stake of this problem for what concerns the primordial nucleosynthesis 
and the galactic evolution of $^3$He is discussed by Tosi and 
by Rood \& Bania in this volume. 
Let us note that in a statistical study of the carbon isotopic ratio 
observed in post-bump RGB stars, Charbonnel \& do Nascimento (1998) showed 
that at least 95 $\%$ of the low mass stars do undergo the extra-mixing. 
This thus leads to a strong revision of the actual contribution of low mass 
stars to $^3$He evolution, and should account for both the measurements 
of $^3$He/H in galactic H$_{\rm II}$ regions (Balser et al. 1994) and in 
the planetary nebulae (Rood et al. 1992, Balser et al. 1997). 

\subsection{The Li rich giants}
If the $^7$Be produced deep inside the RGB stars by $^3$He burning 
could be rapidly transported to cooler regions before
its electron capture to $^7$Li can take place, 
fresh $^7$Li could show up at the stellar surface. 
The so-called Cameron-Fowler (1971) mechanism due to the extra-mixing 
would thus have a chance to produce the so-called Li rich RGB stars 
(de la Reza and Charbonnel \& Balachandran, this volume). 
Sackmann \& Boothroyd (1999; see also Sackmann in this volume and
Denissenkov \& Weiss 2000) showed
that certain assumptions, which depend critically on the speed, geometry 
and episodicity of their parametrized mixing, can indeed lead to important 
Li creation along the RGB. In particular high Li enrichment (up to
logN(Li)=4) is obtained by Sackmann \& Boothroyd (1999) with a continuous 
mixing which simultaneously
induces a smooth decrease of the carbon isotopic ratio up to the RGB
tip. As we discussed in the previous section this prediction for the
$^{12}$C/$^{13}$C ratio is not sustained by the observations 
and cast doubt on the underlying assumptions of the model. 
Moreover Charbonnel \& Balachandran (2000; see also this volume) showed
that the field Li rich stars are not observed all along the RGB, but that they
do clump at the bump phase. This result fits well to the observations of
the $^{12}$C/$^{13}$C ratio which reveal a very fast mixing episode at
the RGB bump (Palacios et al. 2000). 
Thus the Li-rich phase is extremely short, and its
contribution to the Li enrichment of the ISM is certainly negligible.

\subsection{Field-to-cluster differences}
At this point, it is important to remember some striking differences 
which distinguish field and globular cluster red giants. 
In particular, while field giants show C anomalies but no O nor Na variations
(indicating that the extra mixing does not reach the deep internal regions
where the complete CNO cycling and the NeNa cycle occur),
globular cluster bright giants exhibit the so-called universal O-Na anticorrelation 
(e.g., Kraft et al. 1993, Kraft 1994, Denissenkov et al. 1998) which could be 
of primordial origin or due to a deeper extra-mixing than in field stars. 
Up to now, observations in globular clusters are available only for some very 
bright stars close to the RGB tip (except in M13 for which very sparse data 
exist for stars close to the RGB bump) in very few globular clusters. 
Observations are crucially needed in less evolved stars down to the main 
sequence
turnoff in order to quantify the primordial contamination of the intracluster 
gas
by an earlier generation of more massive stars 
and to disentangle it from the real impact of in situ extra-mixing.
In other words, the field-to-cluster differences have to be observationaly 
quantified 
and understood to constrain the physics and the efficiency of the extra-mixing 
in the advanced evolutionary phases.

\section{Acknowledgements}

C.C. thanks the Action Sp\'ecifique de Physique
Stellaire and the Conseil National Fran\c cais d'Astronomie for support.
C.P.D. acknowledges support from the United States
National Science Foundation under grant AST-9812735. 
M.P. would like to acknowledge support from NASA grant NAG5-7150 and
NSF grant AST-9731621.

\end{document}